# Monte Carlo Simulation of X-rays Multiple Refractive Scattering from Fine Structure Objects imaged with the DEI Technique

A. N. Khromova, L. Rigon, F. Arfelli, R. H. Menk, H. J. Besch, H. Plothow-Besch

*Abstract* – In this work we present a novel 3D Monte Carlo photon transport program for simulation of multiple refractive scattering based on the refractive properties of X-rays in highly scattering media, like lung tissue. Multiple scattering reduces not only the quality of the image, but contains also information on the internal structure of the object. This information can be exploited utilizing image modalities such as Diffraction Enhanced Imaging (DEI). To study the effect of multiple scattering a Monte Carlo program was developed that simulates multiple refractive scattering of X-ray photons on monodisperse PMMA (poly-methyl-methacrylate) microspheres representing alveoli in lung tissue.

Eventually, the results of the Monte Carlo program were compared to the measurements taken at the SYRMEP beamline at Elettra (Trieste, Italy) on special phantoms showing a good agreement between both data.

## I. INTRODUCTION

Diffraction Enhanced Imaging (DEI) is a recent phase sensitive technique based on the use of an analyzer crystal placed between the sample and the detector. The original DEI algorithm proposed by Chapman [1] is based on the acquisition of two images, one on each side of the analyzer crystal rocking curve (RC) at 50% of the reflectivity (on the slopes). The images are then processed to obtain the so called *apparent absorption image* and *refraction image*. The sources of contrast in the first are absorption and extinction (scatter rejection). In the refraction image the intensity in each pixel mainly determined by the angle of refraction in the object plane. With such a setup it became possible to improve the visibility of small low absorbing and low contrast structures not visible in the conventional (absorption) X-ray imaging.

Problems arose, however, in case of overlapping microstructures such as alveoli in lung tissue.

Each lung contains around three hundred million alveoli with size from 30 to 100 μm, not detectable separately. Such structures give rise to the effect of multiple refractive scattering. Multiple scattering reduces both the signal-to-noise ratio and the contrast between the healthy and the diseased tissue and contributes to poor diagnostic quality. On the other hand, it could give also additional information on the internal structure of the object.

When the effect of the multiple scattering is present, the signal is proportional to the second derivative of the RC and, therefore, drastically decreases on the half slope regions. On the other hand, a good signal for such fine structure samples is achieved on the top and toes of the RC: 100% and 10% of the reflectivity, respectively. Hence, a new algorithm was proposed [2]. In comparison to the previous method it measures the standard deviation of the refractive angle distribution of X-rays leaving the sample.

Recently other methods have been proposed [3], [4], which are an improvement on the previous DEI algorithms. In these methods multiple images are acquired (Multiple-image radiography MIR) at N different positions of the analyzer RC. It is evident that more accurate results can be obtained using very detailed analysis (pixel-by-pixel) of many more images than in Chapman's original approach. This can be especially useful for the investigation of fine structure samples, since the resolution and the contrast of such objects in DEI is reduced due to the effect of multiple scattering. In MIR these problems are partially resolved. However, deeper investigations are essential to understand in details systems of multiple scatter.

## II. MONTE CARLO SIMULATION PROGRAM

Therefore a Monte Carlo simulation based on a three dimensional vector approach was developed. The proposed algorithm is partially based on the ray tracing algorithm for the three-dimensional graphics [5], [6]. For the ease of the simulation and its experimental verification, lung tissue and subsequently the alveoli are represented by multiple layers of monodisperse PMMA (poly-methyl-methacrylate) microspheres.

This work was supported by the EU under the contract HPRI-CT-1999-50008. The author was individually supported by DAAD grant and INTAS grant under the contract No 03-69-661.

A. N. Khromova is with the Physics department of the Siegen University, Siegen, Germany (e-mail: khromova@deph.physik.uni-siegen.de), on leave from the National University "Kyiv-Mohyla Academy", Kyiv, Ukraine.

L. Rigon is with the INFN and the Physics department of the University Trieste, Trieste, Italy (e-mail: luigi.rigon@ts.infn.it).

F. Arfelli is with the INFN and the Physics department of the University Trieste, Trieste, Italy (e-mail: fulvia.arfelli@ts.infn.it).

R. H. Menk is with the Syncrotrone Trieste, Trieste, Italy (e-mail: ralf.menk@elettra.trieste.it).

H. J. Besch is with the Physics department of the Siegen University, Siegen, Germany (e-mail: besch@deph.physik.uni-siegen.de).

H. Plothow-Besch is with the Physics department of the Siegen University, Siegen, Germany (e-mail: plothow@besch2.physik.uni-siegen.de).

The input parameters of the Monte Carlo program are the spheres' diameter, the X-ray beam energy, the refractive indexes of the spheres' material (PMMA) and the embedding environment (air), X-rays source (synchrotron radiation or X-ray tube radiation) and the thickness of a phantom containing microspheres.

The phantom, simulating lung tissues, is thought to be a flat Plexiglas box in the form of a stair with five steps of different depth: 0.5, 1, 2, 3 and 5 mm filled with monodisperse PMMA (poly-methyl-methacrylate $(C_5H_8O_2)_n$) microspheres. For this simulation two different sizes of microspheres were used: 30 and 100 μm. This sample was exposed to three different x-ray energies of 17 keV, 25 keV and 30 keV.

The phantom used in the experiment and described above has been simulated using SIAMS S3D software [7]. This program builds a close packing of the spheres using an algorithm named 'drop and roll' or 'rolling algorithm', which yields a packing parameter of 0.60. The physical idea of this algorithm is formulated as follows. The spheres are generated according to the chosen law of size distribution (in our case this is a delta function, because all sphere should have the same size) and are dropped inside the box either from one point, or from randomly chosen positions. As soon as the dropped sphere encounters an obstacle (the box wall or an already packed sphere), it sticks to it (without impact) and begins to slide on its surface in the direction of the minimum of the potential energy to the following obstacle. This direction is a projection of the free fall direction on the surface of the obstacle. The movement of the sphere stops, obviously, at a point of intersection of three surfaces (three spheres, two spheres and one plane and etc.) or on a surface situated perpendicular to the direction of the free fall of the sphere (for example, on the bottom of the box).

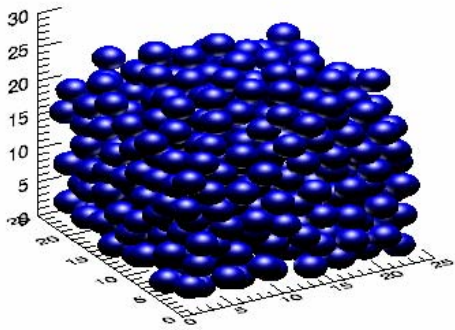

Fig.1. Simulated phantom from the experiment. Diameter of the spheres is 30μm, the box dimensions are 241.8μm ×241.8μm ×241.8μm (for 500 spheres)

The data file, created by this program, contains the Cartesian coordinates of the centers of all spheres with the given radius and the dimensions of the box and it will be included in the Monte Carlo program. In Fig.1 it is possible to see the experimental phantom simulated by the SIAMS 3D software with spheres diameter of 30 μm. The output box dimensions for 500 such a spheres are 241.8μm × 241.8μm × 241.8μm.

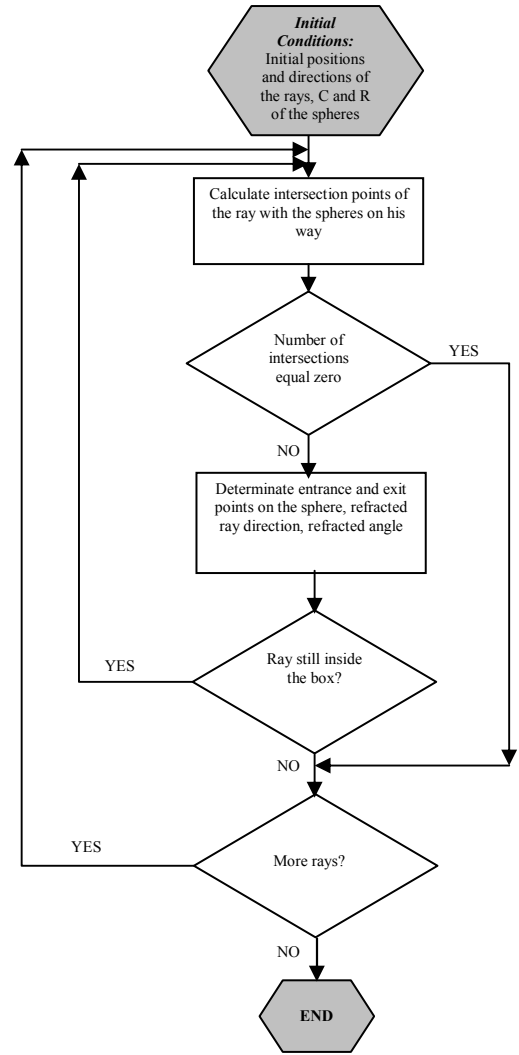

Fig.2. A schematic of the program algorithm

The schematic of the Monte Carlo program algorithm is shown in Fig. 2. C is the position vector of a generic sphere center and R is a radius of this sphere.

We need to investigate the path of each X-ray, its interactions with the microspheres, refracted ray directions and, correspondingly, refraction angles distributions, based upon the refraction properties of X-rays in the media.

Each X-ray is defined by a starting position *S* and a direction vector *d*. The parametric equation of a line is used for the position vector of the ray:

$$P_{ray} = S + \alpha d \qquad (1)$$

The problem now is to find the exact position of the intersection points of the ray with the sphere, if any. The vector *q* will be taken from the centre of the sphere to the point of intersection. This arrangement is shown in the Fig.3.

Vector *q* should be equal to radius ($|q|=r$) to obtain the desired points, which gives a quadratic in α:

$$(S-C) \bullet (S-C) - r^2 + 2\alpha d \bullet (S-C) + \alpha^2 = 0 \quad (2)$$

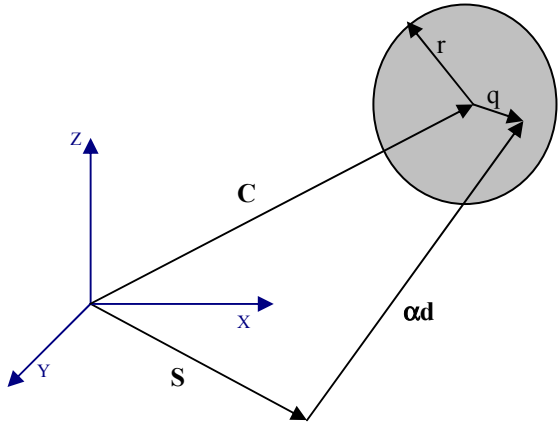

Fig.3. Vectors arrangement for the algorithm calculation

The intersection points are given by the roots of the quadratic equation (2) as:

$$S + \alpha_i d \quad (3)$$
$$i = 1,2$$

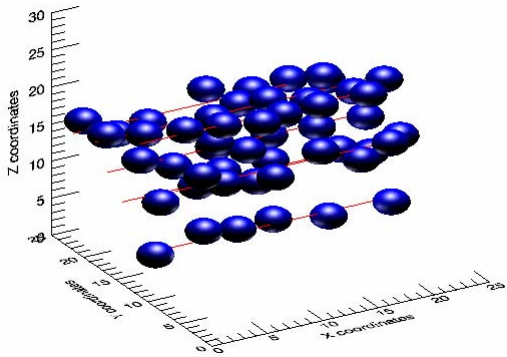

Fig.4. Pass of the X-ray photons inside the box

From this program the refraction angle distributions and their standard deviations are obtained. 100000 photons have been generated for each of the five phantom thicknesses, for each of the three energies (17, 25 and 30 keV) and two spheres diameters (30 and 100 μm). The path of the X-ray photons inside the box can be seen in Fig.4.

*A. Monte Carlo simulation results*

As seen in Fig.5a-5c for 30 μm in diameter spheres at 17 keV, the shape of the refraction angle distribution approaches a Gaussian, as expected. Clearly visible is the broadening of the scattering distribution, while increasing the thickness of the phantom and, thus, the number of scatter. For all other types of the samples an analogue behaviour has been observed.

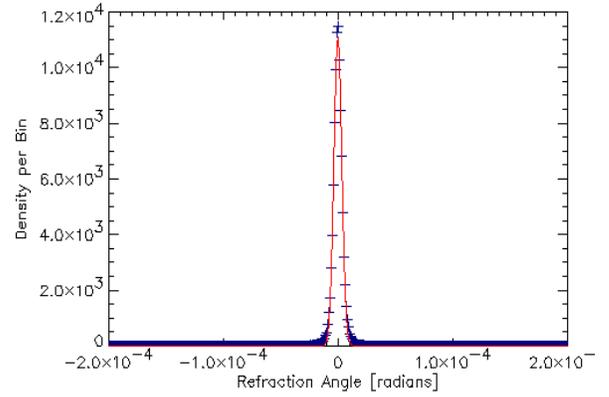

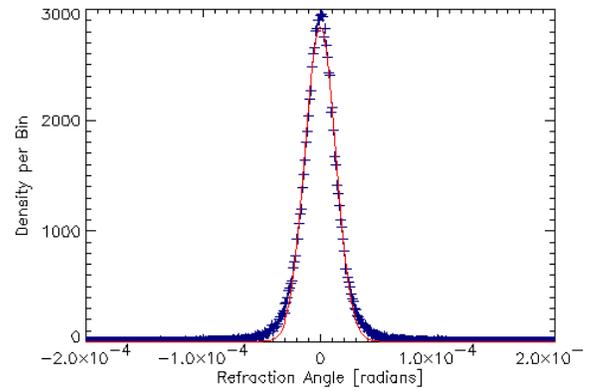

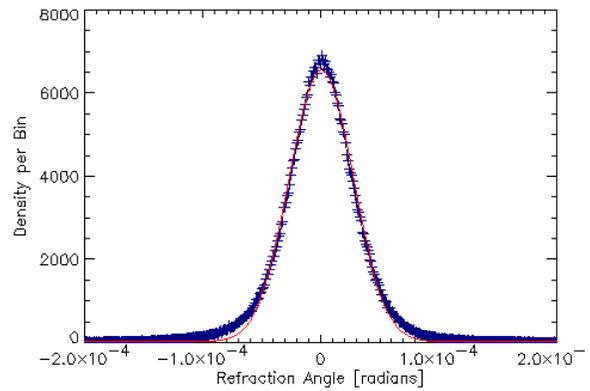

*Fig.5a*: 30μm spheres at 17keV, 0.24 mm phantom thickness (500 spheres), 100000 rays *Fig.5b*: 30μm spheres at 17keV, 0.97 mm phantom thickness (2000 spheres), 100000 rays *Fig.5c*: 30μm spheres at 17keV, 2.9 mm phantom thickness (6000 spheres), 500000 rays

In Fig.6 the standard deviation values versus the phantom thickness for 30 and 100 μm diameter spheres at 17 keV, 25 keV and 30 keV, respectively, are presented. The symbols indicate the simulated values while the lines represent a square root fit to the simulated data, since according to the statistic nature of the multiple scattering process one expect a broadening of the scatter curve that scales with the square root of the number of spheres involved.

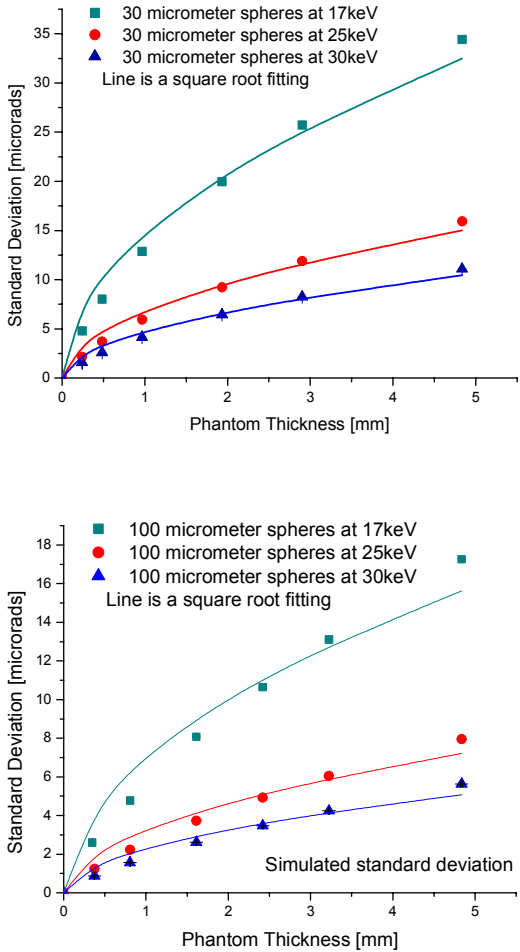

Fig.6. Simulated standard deviation like a square root function of the phantom thickness

The notable discrepancy of the square root fit with the simulated data for small phantom thickness is due to the small amount of scatter involved in the process. In this case a multi scatter approach is not appropriate and a linear behaviour is expected.

From the simulations the energy dependency becomes obvious as well. If the energy is increased from 17 keV to 30 keV, the standard deviation values of the refraction angle distribution decrease for both 30μm and 100μm spheres accordingly. Moreover, an increase of the spheres' diameter results in a decrease of the standard deviation of the refraction angle distribution.

## II. EXPERIMENTS AND RESULTS

Experimental studies on multiple scattering has been carried out at the SYRMEP beamline of the synchrotron radiation facility ELETTRA in Trieste (Italy) utilizing the DEI setup described in detail in references [8], [9]. Phantoms similar to those described before were utilized in multiple image studies using the analyzer crystal. Analysis of the phantom scatter distribution with an analyzer crystal in a multi-image approach results in a broadening of the effective rocking curve. The latter is a convolution of the intrinsic rocking curve with the scatter distribution of the phantom. Thus, deconvolution of the intrinsic rocking curve of the analyzer crystal from the effective rocking curve results in the measured scatter distribution. As depicted in Fig.7a-7b, the standard deviation of the measured scatter distribution is shown as function of the phantom thickness and compared to the simulated data.

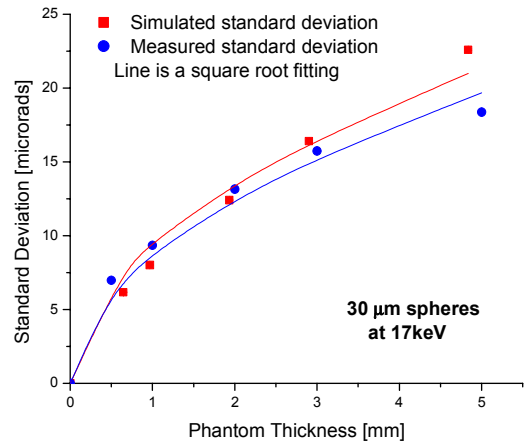

Fig.7a. Comparison between the measured and simulated standard deviation values. 30 μm in diameter spheres at 17 keV

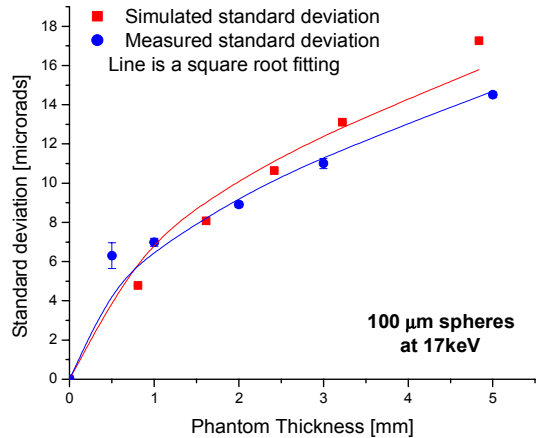

Fig.7b. Comparison between the measured and simulated standard deviation values. 100 μm in diameter spheres at 17 keV

In both cases a square root dependency is notable. The agreement between the measured and simulated results is good if the packing density (0.31 for the 30 μm diameter spheres and

0.6 for the 100 μm diameter spheres) is chosen accordingly in the Monte Carlo simulation program. Of note is that the standard deviation values are strongly dependent on the phantom packing density. The less the packing density of the sample, the smaller are the values of the width of the refraction angle distribution. The larger spheres (for example 100 μm) are usually packed closer to each other due to the influence of the gravitational force, which is dominant compared to the electrostatic one, which occurs during the filling procedure of the PMMA spheres in the Lucite containers. For the smaller spheres electrostatic forces become dominant, which results in a smaller packing density. As in the case of the simulation also in the measured data the notable discrepancy of the square root fit for small phantom thickness is due to the small amount of scatter involved in the process. In this case a multi scatter approach is not appropriate.

### III. SUMMARY AND CONCLUSION

A Monte Carlo program based on a three dimensional vector approach was developed to describe multiple refraction scattering of x-rays in fine objects. Results of the program were experimentally verified utilizing scatter phantoms consisting of microspheres in combination with multiple image diffraction enhanced imaging. A good agreement between simulated and measured data turned out if the packing density is adjusted to 0.6 and 0.31 for the 100 μm and the 30 μm spheres respectively. Such packing densities seem realistic considering the preparation of the phantom. The goal of this study has been to verify the working principle on simple phantoms. In the next step more complex biological samples such as lung tissues can be imitated and examined quantitatively.

### IV. ACKNOWLEDGMENTS

This work was supported by the EU under the contract HPRI-CT-1999-50008. The author was individually supported by the DAAD grant and INTAS grant No 03-69-661.